\documentclass[11pt,onecolumn]{article}
\advance\textheight by \topskip
\usepackage{fixltx2e,fix-cm}
\usepackage{amssymb}
\usepackage{amsmath}
\usepackage{graphicx}
\usepackage{subfigure}
\usepackage{makeidx}
\usepackage{multicol}
\usepackage{lmodern}
\usepackage[T1]{fontenc}

\usepackage{url,color,colortbl}
\usepackage{multirow,booktabs,dcolumn,hhline}
\usepackage{algorithm}
\usepackage[noend]{algpseudocode}
\usepackage{clrs+} 
\renewcommand{\algorithmicrequire}{\textbf{Input:}}
\renewcommand{\algorithmicensure}{\textbf{Output:}}
\renewcommand{\algorithmicindent}{12pt}
\renewcommand{\algorithmicindent}{12pt}

\usepackage{varioref}
\usepackage{t1enc}
\usepackage[latin1]{inputenc}
\usepackage[english]{babel}
\usepackage[square,sort,comma,numbers]{natbib}
\newcommand{\qc}{\c{c}}

\begin{document}
\title{Distributed-Memory Breadth-First Search on Massive Graphs \footnote{To appear as a chapter in \emph{Parallel Graph Algorithms, D. Bader (editor), CRC Press, 2015}}} 

\author{\vspace{0.2cm} Ayd\i n Bulu\qc$^1$, Scott Beamer$^2$, Kamesh Madduri$^3$, Krste Asanovi\'{c}$^2$, David Patterson$^2$ \\
{\em  $^1$Computational Research Division, Lawrence Berkeley National Laboratory, Berkeley, USA} \\
{\em $^2$EECS Department, University of California, Berkeley, USA}\\
{\em $^3$Computer Science \& Engineering Department, The Pennsylvania State University}\\
}
\date{}
\maketitle

\graphicspath{figures}

\newcommand{\vsf}[0]{\vspace{-9pt}} 
\newcommand{\vsb}[0]{\vspace{-6pt}} 
\newcommand{\vectornorm}[1]{\left|\left|#1\right|\right|}
\newcommand{\erdosrenyi}{Erd\H os-R\'{e}nyi}
\newcommand{\transpose}     {^{\mbox{\scriptsize \sf T}}}
\newcommand{\procdecl}[1]   {\proc{#1}\vrule width0pt height0pt depth 7pt \relax}
\newcommand{\lilabel}[1]        {\label{li:#1}}
\newcommand{\liref}[1]      {line~\ref{li:#1}}
\newcommand{\Liref}[1]      {Line~\ref{li:#1}}
\newcommand{\lirefs}[2]     {lines \ref{li:#1}--\ref{li:#2}}
\newcommand{\Lirefs}[2]     {Lines \ref{li:#1}--\ref{li:#2}}
\newcommand{\lireftwo}[2]   {lines \ref{li:#1} and~\ref{li:#2}}
\newcommand{\lirefthree}[3] {lines \ref{li:#1}, \ref{li:#2}, and~\ref{li:#3}}

\def\Cpp{C{}\texttt{++}}
\newcommand{\fix}[1]{\textcolor{red}{\bf #1}}
\newcommand{\rb}[1]{\raisebox{1.5ex}[0pt]{#1}}
\def\naive{na\"\i ve }
\def\Naive{Na\"\i ve }
\def\naively{na\"\i vely }
\def\Naively{Na\"\i vely }

\section{Introduction}\label{intro}

In this chapter, we study the problem of traversing large graphs. A traversal, a systematic method of exploring all the vertices and edges in a graph, can be done in many different orders. A traversal in ``breadth-first'' order, a breadth-first search (BFS), is important because it serves as a building block for many graph algorithms. Parallel graph algorithms increasingly rely on BFS for exploring all vertices in a graph because depth-first search is inherently sequential. Fast parallel graph algorithms often use BFS, even when the optimal sequential algorithm for solving the same problem relies on depth-first search. Strongly connected component decomposition of a graph~\cite{scc_implementation, tarjan_dfs} is an example of such an computation.

Given a distinguished source vertex $s$, BFS systematically explores the graph $G$ to discover every vertex that is reachable from $s$.
In the worst case, BFS has to explore all of the edges in the connected component that $s$ belongs to in order to reach every vertex in the connected component. A simple level-synchronous traversal that explores all of
the outgoing edges of the current frontier (the set of vertices discovered in this level) is therefore considered optimal in the worst-case analysis. This level-synchronous algorithm exposes
lots of parallelism for low-diameter (small-world) graphs~\cite{WS98}. Many real-world graphs, such as those representing social interactions and brain anatomy, are known to have small-world characteristics.

Parallel BFS on distributed-memory systems is challenging due to its low computational intensity and irregular data access patterns. Recently, a large body of optimization strategies have been designed to improve the performance of parallel BFS in distributed-memory systems. Among those, two major techniques stand out in terms of their success and general applicability. The first one is the {\it direction-optimization} by Beamer et al.\cite{Beamer:SC-2012} that optimistically reduces the number of edge examinations by integrating a {\it bottom-up} algorithm into the traversal. The second one is the use of two-dimensional (2D) decomposition of the sparse adjacency matrix of the graph~\cite{sc11-bfs2d, YCH05}.

In this article, we build on our prior distributed-memory BFS work~\cite{beamerbfs13, sc11-bfs2d} by expanding our performance optimizations. We generalize our 2D approach to arbitrary $p_r\times p_c$ rectangular processor grids (which subsumes the 1D cases in its extremes of $p_r=1$ or $p_c=1$). 
We evaluate the effects of in-node multithreading for performance and scalability. 
We run our experiments on three different platforms: a Cray XE6, a Cray XK7 (without using co-processors), and a Cray XC30. We compare the effects of using a scalable hypersparse representation called Doubly Compressed Sparse Column (DCSC) for storing local subgraphs versus a simpler but less memory-efficient Compressed Sparse Row (CSR) representation. Finally, we validate our implementation by using it to efficiently traverse a real-world graph.

\section{Breadth-First Search}
\label{sec:serialtopdown}
Before delving into the details of implementing our parallel algorithm, we review sequential versions of the top-down and bottom-up BFS algorithms.
The level-synchronous top-down BFS can be implemented sequentially using a queue, as shown in Algorithm~\ref{alg:serial-td}. The algorithm outputs an implicit ``breadth-first spanning tree'' rooted 
at $s$ by maintaining parents for each vertex. The parent of a vertex $v$ who is $d$ hops away from the root, can be any of the vertices that are both $d-1$ hops away from the root and 
have an outgoing edge to $v$. This algorithm's running time is proportional to $\Theta(n+m)$ where $n =|V|$ is the number of vertices and $m=|E|$ is the number of edges of a graph $G=(V,E)$. This algorithm's best-case and worst-case performance are equal, since it will always examine the whole connected component the search started from.

The level-synchronous top-down algorithm is overly pessimistic and can be wasteful in practice, because it always performs as many operations as its worst case. Suppose that a vertex $v$ is $d$ hops away from the source and is reachable by $x$ vertices, $x' \leq x$ of which are $d-1$ hops away from the source. In other words, each one of those $x'$ vertices
can potentially be the parent of $v$. In theory, only one of those $x'$ incoming edges of $v$ needs to be explored, but the top-down algorithm is unable to exploit this and does $x'-1$ extra checks. 
By contrast, $v$ would quickly find a parent by checking its incoming edges if a significant number of its neighbors are reachable in $d-1$ of hops of the source. The {\it direction-optimizing} BFS algorithm~\cite{Beamer:SC-2012} 
uses this intuition to significantly outperform the top-down algorithm  because it reduces the number of edge examinations by integrating a {\it bottom-up} algorithm into its search.

The key insight the bottom-up approach leverages is that most edge examinations are unsuccessful because the endpoints have already been visited. In the conventional top-down approach, during each step, every vertex in the frontier examines all of its neighbors and claims the unvisited ones as children and adds them to the next frontier. On a low-diameter graph when the frontier is at its largest, most neighbors of the frontier have already been explored (many of which are within the frontier), but the top-down approach must check every edge in case the neighbor's only legal parent is in the frontier. The bottom-up approach passes this responsibility from the parents to the children (Algorithm~\ref{alg:serial-bu}).

During each step of the bottom-up approach, every unvisited vertex ($parent[u] = -1$) checks its neighbors to see if any of them are in the frontier. If one of them is in the frontier, it is a valid parent and the neighbor examinations (\liref{neighloop1<} -- \liref{neighloop2<}) can end early. This early termination serializes the inner loop in order to get the savings from stopping as soon as a valid parent is found. In general, the bottom-up approach is only advantageous when the frontier constitutes a substantial fraction of the graph. Thus, a high-performance BFS will use the top-down approach for the beginning and end of the search and the bottom-up approach for the middle steps when the frontier is at its largest. Since the BFS for each step is done in whichever direction will require the least work, it is a \emph{direction-optimizing} BFS.

\begin{algorithm}[ht]
\begin{algorithmic}[1]
\Require $G(V,E)$, source vertex $s$
\Ensure $\id{parent}[1..n]$, where $\id{parent}[v]$ gives the parent of $v \in V$ in the BFS tree or $-1$ if it is unreachable from $s$
\State  $\id{parent}[:] \leftarrow -1$, $\id{parent}[s] \leftarrow s$
\State $\id{frontier} \leftarrow \{s\}$, $\id{next} \leftarrow \phi$ 
\While {$\id{frontier} \neq \phi$}
\For {each $u$ in $\id{frontier}$}
\For {each neighbor $v$ of $u$}
\If {$\id{parent}[v] = -1$}
\State  $\id{next} \leftarrow \id{next} \cup \{v\}$
\State	$\id{parent}[v] \leftarrow u$
\EndIf
\EndFor
\EndFor
\State $\id{frontier} \leftarrow \id{next}$, $\id{next} \leftarrow \phi$
\EndWhile
\end{algorithmic}
\caption{Sequential top-down BFS algorithm}
\label{alg:serial-td}
\end{algorithm}


\begin{algorithm}[ht]
\begin{algorithmic}[1]
\Require $G(V,E)$, source vertex $s$
\Ensure $\id{parent}[1..n]$, where $\id{parent}[v]$ gives the parent of $v \in V$ in the BFS tree or $-1$ if it is unreachable from $s$
\State  $\id{parent}[:] \leftarrow -1$, $\id{parent}[s] \leftarrow s$
\State $\id{frontier} \leftarrow \{s\}$, $\id{next} \leftarrow \phi$ 
\While {$\id{frontier} \neq \phi$}
\For {each $u$ in $V$}
\If {$\id{parent}[u] = -1$}
\For {each neighbor $v$ of $u$ \lilabel{neighloop1<}}
\If {$v$ in $\id{frontier}$}
\State  $\id{next} \leftarrow \id{next} \cup \{u\}$
\State	$\id{parent}[u] \leftarrow v$
\State	{\bf break \lilabel{neighloop2<}}
\EndIf
\EndFor
\EndIf
\EndFor
\State $\id{frontier} \leftarrow \id{next}$, $\id{next} \leftarrow \phi$
\EndWhile
\end{algorithmic}
\caption{Sequential bottom-up BFS algorithm}
\label{alg:serial-bu}
\end{algorithm}

\section{Related Work}
In this section, we survey prior work on parallel BFS algorithms and their implementations. A more comprehensive related work on parallel BFS, including results on multithreaded systems and GPUs, can be found in our previous paper~\cite{sc11-bfs2d}.

\subsection{Shared-memory parallel BFS}
Multicore systems have received considerable interest as an evaluation platform due to their prevalence and comparative ease of programming. Current x86 multicore platforms, with 8 to 32-way core-level parallelism and 2-4 way simultaneous multithreading, are much more amenable to coarse-grained load balancing than the prior heavily multithreaded architectures. To alleviate synchronization overheads, successful implementations partition the vertices $p$ ways and replicate high-contention data structures. However, due to the memory-intensive nature of BFS, performance is still quite dependent on the graph size as well as the sizes and memory bandwidths of the various levels of the cache hierarchy.

Recent work on parallelization of the queue-based algorithm by Agarwal et al.~\cite{APPB10} notes a problem with scaling of atomic intrinsics on multi-socket Intel Nehalem systems. To mitigate this, they suggest a partitioning of vertices and corresponding edges among multiple sockets, and a combination of the fine-grained approach and the accumulation-based approach in edge traversal. In specific, the distance values (or the ``visited'' statuses of vertices in their work) of local vertices are updated atomically, while non-local vertices are held back to avoid coherence traffic due to cache line invalidations. They achieve very good scaling going from one to four sockets with this optimization, at the expense of introducing an additional barrier synchronization for each BFS level.

Xia and Prasanna~\cite{XP09} also explore synchronization-reducing optimizations for BFS on Intel Nehalem multicore systems. Their new contribution is a low-overhead ``adaptive barrier'' at the end of each frontier expansion that adjusts the number of threads participating in traversal based on an estimate of work to be performed. They show significant performance improvements over \naive parallel BFS implementations on dual-socket Nehalem systems.

Leiserson and Schardl~\cite{LS10} explore a different optimization: they replace the shared queue with a new ``bag'' data structure which is more amenable for code parallelization with the Cilk++ run-time model. They show that their bag-based implementation also scales well on a dual-socket Nehalem system for selected low diameter benchmark graphs. These three approaches use seemingly independent optimizations and different graph families to evaluate performance on, which makes it difficult to do a head-to-head comparison. Since our target architecture in this study are clusters of multicore nodes, we share some similarities to these approaches.

\subsection{Distributed-memory parallel BFS} 
The general structure of the level-synchronous approach holds in case of distributed memory implementations as well, but fine-grained ``visited'' checks are replaced by edge aggregation-based strategies. With a distributed graph and a distributed array $d$, a processor cannot tell whether a non-local vertex has been previously visited or not. So the common approach taken is to just accumulate all edges corresponding to non-local vertices, and send them to the owner processor at the end of a local traversal. There is thus an additional all-to-all communication step at the end of each frontier expansion. Interprocessor communication is considered a significant performance bottleneck in prior work on distributed graph algorithms~\cite{CDT05, LGH07}. The relative costs of inter-processor communication and local computation depends on the quality of the graph partitioning and the topological characteristics of the interconnection network. As mentioned earlier, the edge aggregation strategy introduces extraneous computation (which becomes much more pronounced in a fully distributed setting), which causes the level-synchronous algorithm to deviate from the $O(m+n)$ work bound.

The BFS implementation of Scarpazza et al.~\cite{SVP08} for the Cell/B.E. processor, while being a multicore implementation, shares a lot of similarities with the general ``explicit partitioning and edge aggregation'' BFS strategy for distributed memory system. The implementation by Yoo et al.~\cite{YCH05} for the BlueGene/L system is a notable distributed memory parallelization. The authors observe that a two-dimensional graph partitioning scheme would limit key collective communication phases of the algorithms to at most $\sqrt{p}$ processors, thus avoiding the expensive all-to-all communication steps. This enables them to scale BFS to process concurrencies as high as 32,000 processors. However, this implementation assumes that the graph families under exploration would have a regular degree distribution, and computes bounds for inter-process communication message buffers based on this assumption. Such large-scale scalability with or without 2D graph decomposition may not be realizable for graphs with skewed degree distributions. Furthermore, the computation time increases dramatically (up to 10-fold) with increasing processor counts, under a weak scaling regime. This implies that the sequential kernels and data structures used in this study are not work-efficient. 

Cong et al.~\cite{CAS10} study the design and implementation of several graph algorithms using the partitioned global address space (PGAS) programming model. PGAS languages and runtime systems hide cumbersome details of message passing-based distributed memory implementations behind a shared memory abstraction, while offering the programmer some control over data locality. Cong's work attempts to bridge the gap between PRAM algorithms and PGAS implementations, again with collective communication optimizations. 
Recently, Edmonds et al.~\cite{edmonds10:hipc-srs} gave the first hybrid-parallel 1D BFS implementation that uses active messages.

Checconi et al.~\cite{IBM:SC-2012} provide a distributed-memory parallelization of BFS for BlueGene/P and BlueGene/Q architectures. They use a very low-latency custom communication layer instead of MPI, and specially optimize it for undirected graphs as is the case for the Graph500 benchmark. Another innovation of their work is to reduce communication by maintaining a prefix sum of assigned vertices, hence to avoid sending multiple parent updates for a single output vertex. In a subsequent publication, the same team extend their work to include the direction optimization idea, albeit using a 1D decomposition~\cite{IBM:IPDPS-2014}.

Satish et al.~\cite{Intel:SC-2012} reduce communication to improve performance of BFS on a cluster of commodity processors. They use a bit vector to communicate the edge traversals, which is not only more compact than a sparse list when the frontier is large, but it also squashes duplicates. They use software pipelining to overlap communication and computation and are even able to communicate information from multiple depths simultaneously. Finally, they use the servers' energy counters to estimate the power consumption of their implementation.

\subsection{Distributed Frameworks with BFS support}
Software systems for large-scale distributed graph algorithm design include the Parallel Boost graph library~\cite{GL05} and the Pregel~\cite{MAB10} framework. Both of these systems adopt a straightforward level-synchronous approach for BFS and related problems. Prior distributed graph algorithms are predominantly designed for ``shared-nothing'' settings. However, current systems offer a significant amount of parallelism within a single processing node, with per-node memory capacities increasing as well. Our paper focuses on graph traversal algorithm design in such a scenario. Powergraph~\cite{gonzalez2012powergraph} advocates a similar
GAS (gather-apply-scatter) abstraction for graph-parallel computations. In a recent independent study~\cite{satishnavigating}, Combinatorial BLAS~\cite{combblas} was found to be the fastest platform in terms of BFS performance, among the ones tested. 
That study evaluated only the top-down algorithm that was available in version 1.3 of Combinatorial BLAS.  

\subsection{Other Parallel BFS Algorithms} 
There are several alternate parallel algorithms to the level-synchronous approach, but we are unaware of any recent, optimized implementations of these algorithms. The fastest-known algorithm (in the PRAM complexity model) for BFS represents the graph as an incidence matrix, and involves repeatedly squaring this matrix, where the element-wise operations are in the min-plus semiring (see~\cite{GM88} for a detailed discussion). This computes the BFS ordering of the vertices in $O(\log n)$ time in the EREW-PRAM model, but requires $O(n^3)$ processors for achieving these bounds. This is perhaps too work-inefficient for traversing large-scale graphs. The level synchronous approach is also clearly inefficient for high-diameter graphs. A PRAM algorithm designed by Ullman and Yannakakis~\cite{UY90}, based on path-limited searches, is a possible alternative on shared-memory systems. However, it is far more complicated than the simple level-synchronous approach, and has not been empirically evaluated. The graph partitioning-based strategies adopted by Ajwani and Meyer~\cite{AM09} in their external memory traversal of high-diameter graphs may possibly lend themselves to efficient in-memory implementations as well.

\section{Parallel BFS}
\subsection{Data Decomposition}
Data distribution plays a critical role in parallelizing BFS on distributed-memory machines. The approach of partitioning vertices to individual processors (along with their outgoing edges) is 
the so-called 1D partitioning. By contrast, 2D partitioning assigns vertices to groups of processors (along with their outgoing edges), which are further assigned to members of the group. 
2D checkerboard partitioning assumes the sparse adjacency matrix of the graph is partitioned as follows:

\begin{equation}
A = \left( 
\begin{array}{c|c|c}
A_{1,1} & \ldots  & A_{1,p_c} \\
\hline
\vdots  & \ddots & \vdots  \\
\hline
A_{p_r,1} & \ldots   & A_{p_r,p_c} 
\end{array} 
\right)
\label{eqn:2dpartitioning}
\end{equation}

Processors are logically organized in a square $p = p_r \times p_c$ mesh, indexed by their row and column indices. 
Submatrix $A_{ij}$ is assigned to processor $P(i,j)$. The nonzeros in the $i$th row of the sparse adjacency matrix 
$A$ represent the outgoing edges of the $i$th vertex of $G$, and 
the nonzeros in the $j$th column of $A$ represent the incoming edges of the $j$th vertex. 
Our top-down algorithm actually operates on the transpose of this matrix in order to maintain the linear algebra abstraction, 
but we will omit the transpose and assume that the input is pre-transposed for the rest of this section.

\subsection{Parallel Top-Down BFS}

\begin{algorithm}[ht]
\begin{algorithmic}[1]
\Require $A$: graph represented by a boolean sparse adjacency matrix, $s$: source vertex id
\Ensure $\pi$: dense vector, where $\pi[v]$ is the predecessor vertex on the shortest path from $s$ to $v$, 
		or $-1$ if $v$ is unreachable
\State $\pi(:) \gets -1$, $\pi(s) \gets s$
\State $f(\id{s}) \gets \id{s} $  \Comment{$f$ is the current frontier}
\For{all processors $P(i,j)$\  \InParallel} 
\While{$f \neq \emptyset$} \lilabel{bfsiter<} 
\State \Call{TransposeVector}{$f_{ij}$}
\State $f_i \gets$ \Call{Allgatherv}{$f_{ij}, P(:,j)$} \lilabel{expand<}
\State $t_{ij}(:) \leftarrow 0$ \Comment{$t$ is candidate parents}
\For {each $f_i(u) \neq 0$}  \Comment{$u$ is in the frontier} \lilabel{seqspmsv<}
\For {each neighbor $v$ of $u$ in $A_{ij}$}
\State  $t_{ij}(v) \leftarrow u$
\EndFor
\EndFor
\State $ t_{ij} \gets $ \Call{Alltoallv}{$t_i, P(i,:)$}  \lilabel{fold<}
\State $f_{ij}(:) \leftarrow 0$
\For {each $t_{ij}(v) \neq 0$} \lilabel{parentupdate<}
\If {$\pi_{ij}(v) \neq -1$} \Comment{Set parent if new}
\State	$\pi_{ij}(v) \leftarrow t_{ij}(v)$
\State $f_{ij}(v) \leftarrow v$ \lilabel{updatepar2<}
\EndIf
\EndFor
\EndWhile
\EndFor
\end{algorithmic}
\caption{Parallel 2D top-down BFS algorithm (adapted from the linear algebraic algorithm~\cite{sc11-bfs2d})}
\label{alg:2dbfs}
\end{algorithm}

The pseudocode for parallel top-down BFS algorithm with 2D partitioning is given in Algorithm~\ref{alg:2dbfs} for completeness. Both $f$ and $t$ are implemented as sparse vectors. 
For distributed vectors, the syntax $v_{ij}$ denotes the local $n/p$ sized piece of the vector owned by the $P(i,j)$th 
processor (not replicated). The syntax $v_i$ denotes the hypothetical $n/p_r$ sized piece of the vector collectively owned by all the processors 
along the $i$th processor row $P(i,:)$ (replicated). 
The algorithm has four major steps:
\begin{itemize}
\item {\bf Expand: } Construct the current frontier of vertices on each processor by a collective allgather step along the processor column (\liref{expand<}). 
\item {\bf Local discovery: } Inspect adjacencies of vertices in the current frontier and locally merge them (\liref{seqspmsv<}). The operation is actually a 
sparse matrix-sparse vector multiplication (SpMSV) on a special semiring where each scalar multiply returns the second operand and each scalar addition returns the minimum.
\item {\bf Fold: } Exchange newly-discovered adjacencies using a collective alltoallv step along the processor row (\liref{fold<}). This step optionally merges updates 
from multiple processors to the same vertex using the first pair entry (the discovered vertex id) as the key.
\item {\bf Local update: } Update distances/parents for unvisited vertices (\liref{parentupdate<}). The new frontier is composed of any entries that was not removed 
from the candidate parents.
\end{itemize}

In contrast to the 1D case, communication in the 2D algorithm happens only along one processor dimension at a time. 
If {\em expand} happens along one processor dimension, then {\em fold} happens along the other processor dimension. 
Both 1D and 2D algorithms can be enhanced by in-node multithreading, resulting in one MPI process per chip instead of one MPI process per core, which will reduce the number of communicating parties.
Large scale experiments of 1D versus 2D show that the 2D approach's communication costs are lower than the respective 1D approach's, with or without in-node multithreading~\cite{sc11-bfs2d}. The study also shows that in-node multithreading gives a further performance boost by decreasing network contention.

\subsection{Parallel Bottom-Up BFS}
\label{sec:bupar}

Implementing a bottom-up BFS on a cluster with distributed memory introduces some challenges that are not present in the shared memory case. The speedup from the algorithm is dependent on fast frontier membership tests and sequentializing the inner loop. 

A high-performance distributed implementation must have fast frontier membership tests which requires it to be able to determine if a vertex is in the frontier without crossing the network. Holding the entire frontier in each processor's memory is clearly unscalable. Fortunately, the 2D decomposition~\cite{sc11-bfs2d,YCH05} greatly aids this, since for each processor, only a small subset of vertices can be the sources of a processor's incoming edges. This subset is small enough that it can fit in a processor's memory, and the frontier can be represented with a dense vector for constant time access. The dense format, especially when compressed by using a bitmap, does not necessarily consume more memory than a sparse vector if there are enough non-zeros (active vertices in the frontier). During the bottom-up steps, the frontier should be a large fraction of the graph, so the bottom-up format will be advantageous.

Although the 2D decomposition helps with providing fast frontier checks, it complicates sequentializing the inner loop. Since all of the edges for a given vertex are spread across multiple processors, the examination of a vertex's neighbors will be done in parallel. If the inner loop is not serialized, the bottom-up approach's advantage of terminating the inner loop early once a parent is found will be hard to maintain. Unnecessary edges could be examined during the time it takes for the termination message to propagate across the network.

\begin{figure}
\centering
\includegraphics[width=0.55 \columnwidth]{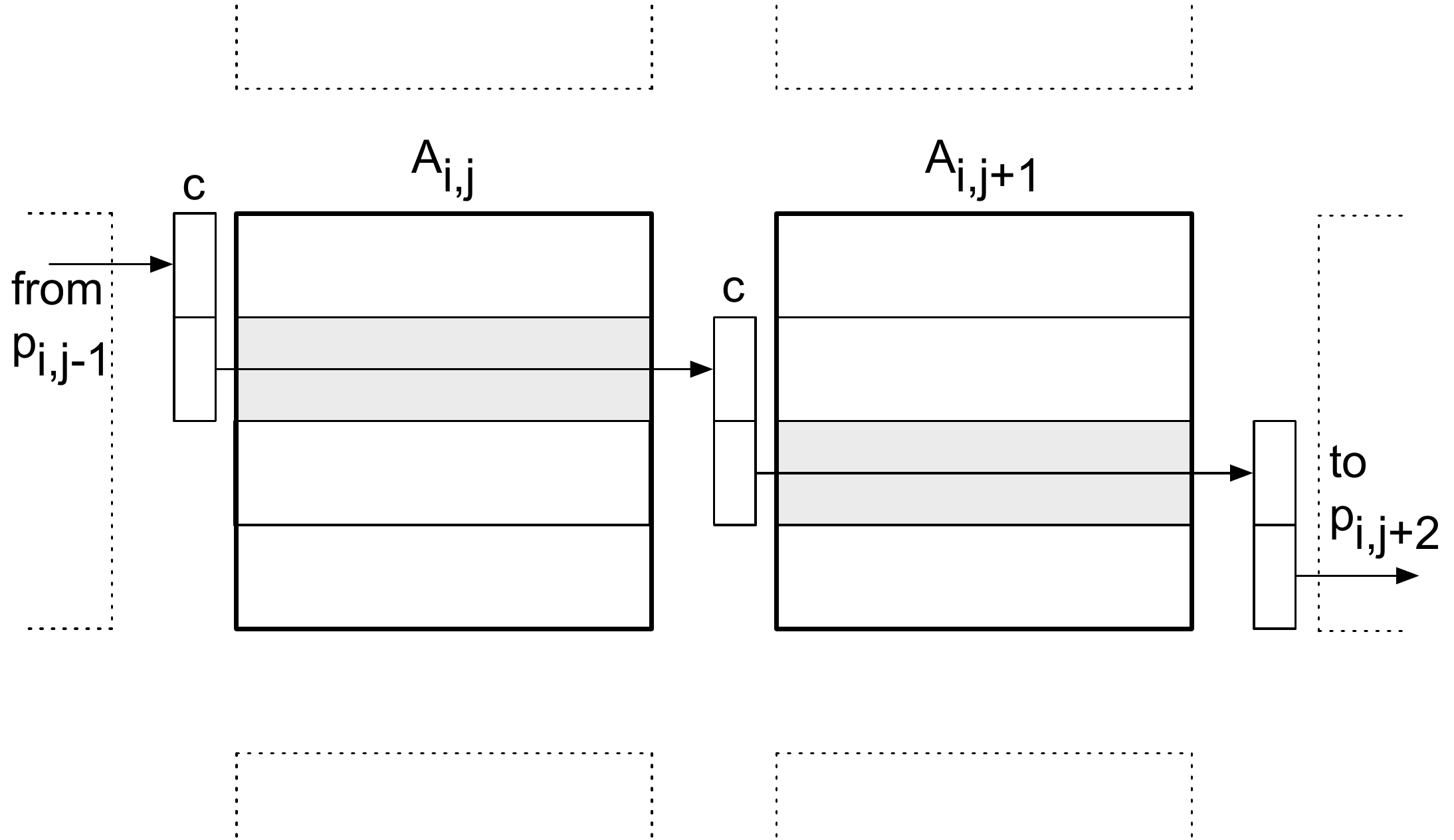}
\caption{Sub-step for processors $p_{i,j}$ and $p_{i,j+1}$.  They initially use their segment of {\it completed} (c) to filter which vertices to process from the shaded region and update {\it completed} for each discovery.  At the end of the sub-step, the {\it completed} segments rotate to the right. The parent updates are also transmitted at the end of the sub-step (not shown).}
\label{figure:rotate}
\end{figure}

To serialize the inner loop of checking if neighbors are in the frontier, we partition the work temporally (Figure~\ref{figure:rotate}). For each BFS level, we break down the search step into $p_c$ sub-steps during which each vertex's edges will be examined by only one processor. During each sub-step, a processor processes $(1/p_c)$th of the vertices in that processor row. After each sub-step, it passes on the responsibility for those vertices to the processor to its right and accepts new vertices from the processor to its left. This pairwise communication sends which vertices have been completed (found parents), so the next processor knows to skip over them. This has the effect of the processor responsible for processing a vertex rotating right along the row each sub-step. When a vertex finds a valid parent to become visited, its index along with its discovered parent is queued up and sent to the processor responsible for the corresponding segment of the parent array to update it.

\begin{algorithm}[ht]
\begin{algorithmic}[1]
\Require $A$: unweighted graph represented by a boolean sparse adjacency matrix, $s$: source vertex id
\Ensure $\pi$: dense vector, where $\pi[v]$ is the parent vertex on the shortest path from $s$ to $v$, or $-1$ if $v$ is unreachable
\State $f(:) \leftarrow 0$, $f(s) \leftarrow 1$ \Comment{bitmap for frontier}
\State $c(:) \gets 0$, $c(\id{s}) \gets 1$ \Comment{bitmap for completed}
\State $\pi(:) \gets -1$, $\pi(s) \gets s$
\While{$f(:) \neq 0$}
\For{all processors $P(i,j)$\  \InParallel}
\State \Call{TransposeVector}{$f_{ij}$} \lilabel{distfront1<}
\State $f_i \gets$ \Call{Allgatherv}{$f_{ij}, P(:,j)$} \lilabel{distfront2<}
\For{s in $0 \dots p_c-1$} \Comment{$p_c$ sub-steps}
\State	$t_{ij}(:) \leftarrow 0$ \Comment{$t$ holds parent updates}
\For{u in $V_{i,j+s}$} \lilabel{localsearch1<}
\If {$c_{ij}(u) = 0$} \Comment{$u$ is unvisited}
\For {each neighbor $v$ of $u$ in $A_{ij}$}
\If {$f_i(v) = 1$}
\State  $t_{ij}(u) \leftarrow v$
\State $c_{ij}(u) \leftarrow 1$
\State {\bf break \lilabel{localsearch2<}}
\EndIf
\EndFor
\EndIf
\EndFor
\State  $f_{ij}(:) \leftarrow 0$ \lilabel{updatepar1<}
\State $w_{ij} \gets$ \Call{Sendrecv}{$t_{ij}, P(i,j+s), P(i,j-s))$}
\For {each $w_{ij}(u) \neq 0$}
\State	$\pi_{ij}(u) \leftarrow w_{ij}(u)$
\State $f_{ij}(u) \leftarrow 1$ \lilabel{updateparbu2<}
\EndFor
\State $c_{ij} \gets$ \Call{Sendrecv}{$c_{ij}, P(i,j+1), P(i,j-1))$} \lilabel{rotate<}
\EndFor
\EndFor
\EndWhile
\end{algorithmic}
\caption{Parallel 2D bottom-up BFS algorithm}
\label{alg:par-bu}
\end{algorithm}

%
%

The pseudocode for our parallel bottom-up BFS algorithm with 2D partitioning is given in Algorithm~\ref{alg:par-bu} for completeness. $f$ ($\id{frontier}$) is implemented as a dense bitmap and $\pi$ ($\id{parents}$) is implemented as a dense vector of vertex identifiers (integers). $c$ ($\id{completed}$) is a dense bitmap and it represents which vertices have found parents and thus no longer need to search. The temporaries $t$ and $w$ are implemented as sparse vectors, but they can be thought of as queues of updates represented as pairs of vertices of the form (child, parent). All processor column indices are modulo $p_c$ (the number of processor columns). For distributed vectors, the syntax $f_{ij}$ denotes the local $n/p$ sized piece of the frontier owned by the $P(i,j)$th processor. Likewise, the syntax $V_{i,j}$ represents the vertices owned by the $P(i,j)$th processor. The syntax $f_j$ denotes the hypothetical $n/p_c$ sized piece of the frontier collectively owned by all the processors along the $j$th processor column $P(:,j)$.
Each step of the algorithm has four major operations:
\begin{itemize}
\item {\bf Gather frontier} (per step) Each processor is given the segment of the frontier corresponding to their incoming edges (\lireftwo{distfront1<}{distfront2<}).
\item {\bf Local discovery} (per sub-step) Search for parents with the information available locally (\liref{localsearch1<} -- \liref{localsearch2<}).
\item {\bf Update parents} (per sub-step) Send updates of children that found parents and process updates for own segment of $parents$ (\liref{updatepar1<} -- \liref{updateparbu2<}).
\item {\bf Rotate along row} (per sub-step) Send $\id{completed}$ to right neighbor and receive $\id{completed}$ for the next sub-step from left neighbor (\liref{rotate<}).
\end{itemize}

\subsection{Parallel Direction-Optimizing BFS}\label{sec:dirbfs}
We implement a parallel direction-optimizing BFS by combining our parallel top-down implementation with our parallel bottom-up implementation. For each BFS level (depth), we choose whichever approach will be faster. The bottom-up approach is faster when the frontier is large since it is more likely to find valid parents sooner and thus skip edges. When the frontier is small, we use the top-down approach because the bottom-up approach will be slower as it will struggle to find valid parents and thus do redundant work. To choose which approach to use, we use simple heuristics from prior work~\cite{beamerTR,beamerbfs13} of using the number of edges originating in the frontier while top-down and using the size of the frontier while bottom-up.

\section{Implementation Considerations}
\label{sec:datastructures}

\subsection{Graph Representation}

We analyze two different ways of representing subgraphs locally in each processor. Our first approach uses a `compressed sparse row' (CSR)-like representation for storing adjacencies. All adjacencies of a vertex are sorted and compactly stored in a contiguous chunk of memory, with adjacencies of vertex $i+1$ next to the adjacencies of $i$. For directed graphs, we store only edges going out of vertices. Each edge (rather the adjacency) is stored twice in case of undirected graphs. An array of size $n+1$ stores the start of each contiguous vertex adjacency block. We use 64-bit integers to represent vertex identifiers. 

The CSR representation is extremely fast because it provides constant-time access to vertex adjacencies. However, CSR is space-efficient only for $p_c=1$ (1D case). We say that a distributed data structure is {\em space efficient} if and only if the aggregate storage for the distributed data structure is on the same order as the storage that would be needed to store the same data structure serially on a machine with large enough memory. A CSR-like representation is asymptotically suboptimal for storing sub-matrices after 2D partitioning for $p_c > 1$. The aggregate memory required to locally store each submatrix in 
CSR format is $O(n \cdot p_c + m)$, while storing the whole matrix in CSR format would only take $O(n + m)$. 
Consequently, a strictly $O(m)$ data structure with fast indexing support is required. 
The indexing requirement stems from the need to provide near constant time access to individual rows (or columns) 
during the SpMSV operation. One such data structure, doubly-compressed sparse columns (DCSC), has been introduced 
before~\cite{ipdps08} for hypersparse matrices that arise after 2D decomposition. DCSC for BFS consists of an array $\mathsf{IR}$ of row ids (size $m$), which is indexed by two parallel arrays of column pointers ($\mathsf{CP}$)
and column ids ($\mathsf{JC}$). The size of these parallel arrays are on the order of the number of columns that has at least one nonzero ($\id{nzc}$) in them. 

For the multithreaded 2D algorithm, we split the node local matrix rowwise to $t$ pieces, as shown 
in Figure~\ref{fig:nonzerostructure} for two threads.  Each thread local $n/(p_r t) \times n/p_c$ sparse matrix is stored in 
DCSC format.

\begin{figure}
\begin{center}
$A =   \left(   
		\begin{array}{c|cccccc}
			& 0 		& 1		& 2		&3 		& 4		& 5  		\\
		\hline
		0	& \times	&    		&  		& \times	& 		&   		\\
		1	& \times 	&  \times   	&  		& 		& 		&   		\\
		2	&  		&     		& 	 	& \times	& 		& \times  	\\
		\hline
		3	& \times 	&     		&  \times	& \times	& 		&   		\\
		4	&  		&  \times   	&  		& 		& 		&   		\\
		5	&  		&     		& \times 	& 		& 		&   			
	        \end{array}\right) $
\end{center}
\caption{Nonzero structure of node-local matrix $A$
\label{fig:nonzerostructure}}
\end{figure}

A compact representation of the frontier vector is also important. It should be represented in a sparse format, where only the indices of the non-zeros are stored. We use a queue in the 1D implementation and a sorted sparse vector in the 2D implementation. Any extra data that are piggybacked to the frontier vectors adversely affect the performance, since the communication volume of the BFS benchmark is directly 
proportional to the size of this vector. 

\subsection{Local Computation}

{\bf Top-Down Computation}:
The node-local multithreaded implementation of the distributed CSR-based BFS is similar to the local computation in Algorithm~\ref{alg:2dbfs}. It maintains a queue of newly-visited vertices. Since a shared queue for all threads would involve thread contention for every insertion, we use thread-local queues for storing these vertices, and merging them at the end of each iteration to form the frontier queue. Next, the distance checks and updates are typically made atomic to ensure that a new vertex is added only once to the queue. However, the BFS algorithm is still correct even if a vertex is added multiple times, as the distance value is guaranteed to be written correctly after the thread barrier and memory fence at the end of a level of exploration. Cache coherence further ensures that the correct value propagates to other cores once updated. We observe that we actually perform a very small percentage of additional insertions (less than $0.5$\%) for all the graphs we experimented with at six-way threading. This lets us avert the issue of non-scaling atomics across multi-socket configurations~\cite{APPB10}. This optimization was also considered by Leiserson et al.~\cite{LS10} (termed ``benign races'') for insertions to their bag data structure.

For the DCSC-based implementation, the local computation time is dominated by the sequential SpMSV operation in \liref{seqspmsv<} of Algorithm~\ref{alg:2dbfs}. This corresponds to selection, scaling and finally merging columns of the local adjacency matrix that are indexed by the nonzeros in the sparse vector. Computationally, we form the union $\bigcup{A_{ij}(:,k)}$ for all $k$ where $f_i(k)$ exists.

We explored multiple methods of forming this union. The first option is to use a priority-queue of size $\id{nnz}(f_i)$ and perform a unbalanced multiway merging. 
While this option has the advantage of being memory-efficient and automatically creating a sorted output, the extra logarithmic factor hurts the performance at small concurrencies, even after using a highly optimized cache-efficient heap. The cumulative requirement for these heaps are $O(m)$. The second option is to use a sparse accumulator (SPA)~\cite{smatlab} which is composed of a dense vector of 
values, a bit mask representing the ``occupied'' flags, and a list that keeps the indices of existing elements in the output vector. The SPA approach proved to be faster in many cases,  although it has disadvantages such as increasing the memory footprint due to the temporary dense vectors, and having to explicitly sort the indices at the end of the 
iteration. The cumulative requirement for the SPAs are $O(n \, p_c)$. We use the SPA approach in this work due to its faster overall performance.

{\bf Bottom-Up Computation}:
The local parent discovery for the bottom-up computation implemented in \lirefs{localsearch1<}{localsearch2<} of Algorithm~\ref{alg:par-bu}.
The most important data structures are the dense bitmap $f$ that holds the current frontier and the dense 
bitmap $c$ that represents vertices that have found parents. 
On a single compute node, a bitmap that fits in the last level of cache provides fast (constant time) membership tests for the frontier. 
Sequentializing the inner loop is trivial since the outer loop can still provide sufficient parallelism to achieve good multicore performance.

\section{Communication Analysis}
\label{sec:anal}
{\bf Top-Down}:
We use a simple linear model to capture the cost of regular (unit stride or fixed-stride) and irregular memory references to various levels of the memory hierarchy, as well as to succinctly express inter-processor MPI communication costs. We use the terms $\alpha$ and $\beta$ to account for the latency of memory accesses and the transfer time per memory word (i.e., inverse of bandwidth) respectively. Further, we use $\alpha_{L}$ to indicate memory references to local memory, and $\alpha_{N}$ to denote message latency over the network (remote memory accesses). The bandwidth terms can also be similarly defined. To account for the differing access times to various levels of the memory hierarchy, we additionally qualify the $\alpha$ term to indicate the size of the data structure (in memory words) that is being accessed. $\alpha_{L,x}$, for instance, would indicate the latency of memory access to a memory word in a logically-contiguous memory chunk of size $x$ words. Similarly, to differentiate between various inter-node collective patterns and algorithms, we qualify the network bandwidth terms with the communication pattern. For instance, $\beta_{N, p2p}$ would indicate the sustained memory bandwidth for point-to-point communication, $\beta_{N, a2a}$ would indicate the sustained memory bandwidth per node in case of an all-to-all communication scenario, and $\beta_{N, ag}$ would indicate the sustained memory bandwidth per node for an allgather operation. 

Using synthetic benchmarks, the values of $\alpha$ and $\beta$ defined above can be calculated offline for a particular parallel system and software configuration. The programming model employed, the messaging implementation used, the compiler optimizations employed are some software factors that determine the various $\alpha$ and $\beta$ values.

For the 2D top-down algorithm (Algorithm~\ref{alg:2dbfs}), consider the general 2D case processor grid of $p_r \times p_c$. The size of the local adjacency matrix is $n/p_r \times n/p_c$. The number of memory
references is the same as the 1D case, cumulatively over all processors. However, the cache working set is bigger, because 
the sizes of the local input (frontier) and output vectors are $n/p_r$ and $n/p_c$, respectively. 
The local memory reference cost is given by 
\[\frac{m}{p}\beta_{L} + \frac{n}{p}\alpha_{L,n/p_c} + \frac{m}{p}\alpha_{L,n/p_r}\] 
The higher number of cache misses associated with larger working sets is perhaps the primary reason for the relatively higher 
computation costs of the 2D algorithm.  

Most of the costs due to remote memory accesses is concentrated in two operations. The expand phase is characterized by an Allgatherv 
operation over the processor column of size $p_r$ (\liref{expand<}) and the fold phase is characterized by an
Alltoallv operation over the processor row of size $p_c$ (\liref{fold<}). 

The aggregate input to the expand (Allgatherv) step is $O(n)$ over all iterations. 
However, each processor receives a $1/p_c$ portion of it, meaning that frontier subvector gets replicated along the processor column.
Hence, the per node communication cost is 
\[p_r\alpha_{N} + \frac{n}{p_c}\beta_{N, ag}(p_r)\]

The aggregate input to the fold (Alltoallv) step can be as high as $O(m)$, although the number is
lower in practice due to in-node aggregation of newly discovered vertices that takes
place before the communication. Since each processor receives only a $1/p$ portion of this data,
the remote costs due to this step are at most 
\[p_c\alpha_{N} + \frac{m}{p}\beta_{N, a2a}(p_c)\]

Our analysis successfully captures that the relatively lower communication costs of the 2D algorithm by
representing $\beta_{N, x}$ as a function of the processor count.

{\bf Bottom-Up}:
For the 2D bottom-up algorithm (Algorithm~\ref{alg:par-bu}) we model only the interprocessor communication since it is the probable bottleneck. We present a simple model that counts the number of 64-bit words sent and received during the entire search. We use 64-bit words as the unit because that is the size of a vertex identifier, which enables us to scale to graphs with greater than $2^{32}$ vertices. When representing the compression provided by bitmaps, we divide the number of elements by 64. To further simplify the expressions, we assume $(p_c-1)/(p_c) \approx 1$ and ignore transfers that send only a word (communicating sizes). We calculate the data volume for the entire search, and assume that every vertex and every edge is part of the connected component.



\begin{table}
\begin{center}
\begin{tabular}{l|c|c|c}
Operation & Type & $\frac{\text{Communications}}{\text{Step}}$ & $\frac{\text{64-bit Words}}{\text{Search}}$ \\
\hline
Transpose & p2p & $O(1)$ & $s_bn/64$ \\
Frontier Gather & ag & $O(1)$ & $s_bnp_r/64$ \\
Parent Updates & p2p & $O(p_c)$ & $2n$ \\
Rotate Completed & p2p & $O(p_c)$ & $s_bnp_c/64$ \\
\hline
\multicolumn{1}{l}{Total} &  & $O(p_c)$ & $n(\frac{s_b(p_r + p_c + 1)}{64} + 2)$
\end{tabular}
\caption{Bottom-up Communication Costs}
\label{table:bu-model}
\end{center}
\end{table}

From our previous analysis, the aggregate communication volume over all processors for the top-down steps is 
\[w_t = p \, O( \frac{n}{p_c} + \frac{m}{p})\]
Going beyond the $O$-notation, we see that communicating an edge means sending both endpoints (two words). 
Since the graph is undirected, each edge is examined from both sides, which results in sending $4m$ words. Consequently, 
the number of words a search with  the top-down approach sends is approximately: 
\[w_t = 4m + n \, p_r\]

Since the bottom-up approach is most useful when combined with the top-down approach, we assume the bottom-up approach is used for only $s_b$ steps of the search, but it still processes the entire graph. There are three types of communication that make up the bottom-up approach: gathering the frontier, communicating completed vertices, and sending parent updates. Gathering the frontier is the same combination of a transpose and an allgather along a column like the top-down approach except a dense bitmap is used instead of a sparse vector. Since the bottom-up approach uses a dense data structure and it sends the bitmap every step it is run, it sends $s_b n(1+p_r)/64$ words to gather the frontier. To rotate the bitmaps for $\id{completed}$, it transfers the state of every vertex once per sub-step, and since there are $p_c$ sub-steps, an entire search sends $s_b n p_c/64$ words. Each parent update consists of a pair of words (child, parent), so in total sending the parent updates requires $2n$ words. All combined, the number of words the bottom-up approach sends is approximately:
\[w_b = n(\frac{s_b(p_r + p_c + 1)}{64} + 2)\]

To see the reduction in data volume, we take the ratio of the number of words the top-down approach sends ($w_t$) to the number of words the bottom-up approach will send ($w_b$), as shown in Equation~\ref{eq:ratio}. We assume our 2D partitioning is square ($p_r = p_c$) since that will send the least amount of data for both approaches. Furthermore, we assume the degree of the target graph is $k = m/n$.
\begin{equation}
\label{eq:ratio}
\frac{w_t}{w_b} = \frac{p_c + 4k}{s_b(2p_c + 1)/64 + 2}
\end{equation}

For a typical value of $s_b$ (3 or 4), by inspection the ratio will always be greater than 1; implying the bottom-up approach sends less data. Both approaches suffer when scaling up the number of processors, since it increases the communication volume. This is not unique to either approach presented, and this leads to sub-linear speedups for distributed BFS implementations. This ratio also demonstrates that the higher the degree is, the larger the gain is for the bottom-up approach relative to the top-down approach. Substituting typical values ($k=16$ and $p_c=128$), the bottom-up approach needs to take $s_b \approx 47.6$ steps before it sends as much data as the top-down approach. A typical $s_b$ for the low-diameter graphs examined in this work is 3 or 4, so the bottom-up approach typically moves an order of magnitude less data. This is intuitive, since to first order, the amount of data the top-down approach sends is proportional to the number of edges, while for the bottom-up approach, it is proportional to the number of vertices.

The critical path of communication is also important to consider. The bottom-up approach sends less data, but it could be potentially bottlenecked by latency. Each step of the top-down algorithm has a constant number of communication rounds, but each step of the bottom-up approach has $\Theta(p_c)$ rounds which could be significant depending on the network latencies.

The types of communication primitives used is another important factor since primitives with more communicating parties may have higher synchronization penalties. This is summarized in Table~\ref{table:bu-model} with the abbreviations: p2p=point-to-point, ag=allgather, and a2a=all-to-all. The communication primitives used by top-down involve more participants, as it uses: point-to-point (transpose to set up expand), allgather along columns (expand), and all-to-all along rows (fold). The bottom-up approach uses point-to-point for all communication except for the allgather along columns for gathering the frontier.

\section{Experimental Design}
\label{sec:expr}

\subsection{Evaluation Platforms}
We evaluate our algorithms on three supercomputers: Cray XE6 at NERSC (Hopper)~\cite{Hopper_website}, Cray XC30 at NERSC (Edison)~\cite{Edison_website}, and Cray XK6 at ORNL (Titan)~\cite{Titan_website}. 
Architectural details of these computers are listed in Table~\ref{tab:machines}. In our experiments, we ran only on the CPUs and did not utilize Titan's GPU accelerators.

In all three supercomputers, we used Cray's MPI implementation, which
is based on MPICH2. All three chip architectures achieve memory parallelism via hardware prefetching. On Hopper, we compiled our code with GCC \Cpp\ compiler version 4.8.1 with \texttt{\mbox{\small -O2 -fopenmp}} flags.
On Titan, we compiled our code using GCC \Cpp\ compiler version 4.6.2 with \texttt{\mbox{\small -O2 -fopenmp}} flags.  
On Edison, we compiled our code using the Intel \Cpp\ compiler (version 14.0.2) with the options \texttt{\mbox{\small -O2 -no-ipo -openmp}}.

{
\setlength{\tabcolsep}{5pt}
\begin{table}[!tb]{
\centering
\begin{tabular}{rccc}
					& {\bf Cray XE6} & {\bf Cray XK7 } & {\bf Cray XC30 }    \\
					& {\bf (Hopper)} & {\bf (Titan)}	& {\bf (Edison)} \\
\hline
\multirow{2}{*}{{\bf Core}} & {\bf AMD}  & {\bf AMD}  & {\bf Intel}		\\
				         & {\bf Magny-Cours} & {\bf Interlagos} & {\bf Ivy Bridge} \\
\hline
Clock (GHz)			& 2.1	 		& 2.2				& 2.4					\\
Private Cache (KB)		& 64+512 		& 16+2048 		& 32+256				\\
DP GFlop/s/core		& 8.4 		& 8.8				&19.2		\\
\hline
\multirow{2}{*}{{\bf Chip Arch.}}	& {\bf Opteron} & {\bf Opteron} & {\bf Xeon}	\\
					& {\bf 6172}	& {\bf 6274}	& {\bf E5-2695 v2} \\
\hline
Cores per chip			& 6				& 16				& 12					\\
Threads per chip		& 6				& 16				& 24$^1$				\\
L3 cache per chip		& 6~MB 			& 2$\times$8~MB	&	30~MB			\\
\hline
{\bf Node Arch.}	&  Hypertransport & 	Hypertransport 		& QPI (8 GT/s)\\
\hline
Chips/node			& 4			&  1		&	2					\\
STREAM BW$^2$	& 49~GB/s 	&  52~GB/s 	&	104~GB/s		\\
Memory per node		& 32~GB		&  22~GB	&	64~GB			\\
\hline
\multirow{2}{*}{Interconnect}		& Gemini & Gemini & Aries 	\\
					&  (3D Torus) &  (3D Torus) & (Dragonfly)	\\
\hline
\end{tabular}
\caption{Overview of Evaluated Platforms.  $^1$Only 12 threads were used.  $^2$Memory bandwidth is measured using the STREAM copy benchmark per node.}
\label{tab:machines}
}
\end{table}
}

\subsection{Data Sets}
We use synthetic graphs based on the R-MAT random graph model~\cite{CZF04}, as well as the largest publicly available real-world graph (at the time of our experiments) that represents the structure of the Twitter social network~\cite{Twitter}, which has 61.5 million vertices and 1.47 billion edges. The Twitter graph is anonymized to respect privacy.  R-MAT is a recursive graph generator that creates networks with skewed degree distributions and a very low graph diameter. R-MAT graphs make for interesting test instances because traversal load-balancing is non-trivial due to the skewed degree distribution and the lack of good graph separators. Common vertex relabeling strategies are also expected to have a minimal effect on cache performance. We use undirected graphs for all of our experiments, even though the algorithms presented
in this chapter are applicable to directed graphs as well.

We set the R-MAT parameters $a$, $b$, $c$, and $d$ to $0.57,0.19,0.19,0.05$ respectively and set the degree to $16$ unless otherwise stated. These parameters are identical to the ones used for generating synthetic instances in the Graph500 BFS benchmark~\cite{Graph500}. Like Graph500, to compactly describe the size of a graph, we use the \emph{scale} variable to indicate the graph has $2^{scale}$ vertices.

When reporting numbers, we use the performance rate TEPS, which stands for Traversed Edges Per Second. Since the bottom-up approach may skip many edges, we compute the TEPS performance measure consistently by dividing the number of input edges by the runtime. During preprocessing,
we prune duplicate edges and vertices with no edges from the graph. For all of our timings, we do 16 to 64 BFS runs from randomly selected distinct starting vertices and report the harmonic mean.

\subsection{Implementation and Versions Evaluated}
\label{sec:distimpl}

We use the MPI message-passing library to express the inter-node communication steps. In particular, we extensively utilize the collective communication routines Alltoallv, Allreduce, and Allgatherv.

Most of our results use DCSC as its main data structure where we rely on the linear-algebraic primitives of the Combinatorial BLAS 
framework~\cite{combblas}, with certain BFS specific optimizations enabled. In particular, we implement the bottom-up traversal 
(and hence the direction-optimizing algorithm)
only using DCSC on Combinatorial BLAS. 

We, however, include comparisons with a stand alone implementation that uses the more conventional CSR data structure. 
CSR offers the benefit of fast indexing to vertex adjacencies, but comes with memory scaling limitations.  
We also use our CSR-based implementation when experimenting with different (rectangular) grid topologies. 

We report our timings using the default MPI rank ordering but our runs occasionally perform better using a different ordering. 
Since collective communication in our 2D algorithms take place on subcommunicators, 
the assignment of MPI tasks to compute nodes affects the performance of the overall algorithm. 
We have the freedom to perform either one of the communication stages (Allgatherv and Alltoallv) in contiguous ranks where processes in the same 
subcommunicator map to sockets that are physically close to each other. The effect of changing default task to processor mapping is not conclusive.
We observed a modest performance increase of up to 20\% for medium concurrencies (10-40k cores) but the benefits are less pronounced in small 
(less than 10k cores) and large concurrencies (over 40k cores). Consequently, we decided to only report results using the default mapping. 
A more detailed study on the effects of MPI task mapping to the collectives used in parallel BFS can be found in earlier work~\cite{dimacs12}. 
While DCSC based implementation performs Alltoallv along processor rows and Allgatherv along processor columns, our CSR based implementation
performs the opposite, i.e., Alltoallv along processor columns and Allgatherv along processor rows.

\section{Experimental Results}
As discussed previously, there are many different ways to implement parallel BFS on a distributed memory architecture. One can use the classical top-down algorithm or the direction-optimizing algorithm. 
One can use a standard CSR implementation for local computation or more memory scalable data structures such as DCSC. One can take advantage of multithreading. 
One can change the process grid dimensions from a perfect square to more rectangular shapes. All in all, it is almost infeasible to experiment with all possible combinations.
In this chapter, we typically change one variable and keep others constant (i.e., {\em ceteris paribus}) to see the effect of that variable on the performance achieved. 

\subsection{Top-down vs. Direction-Optimizing BFS}
\begin{figure}
\centering
\includegraphics[width=0.50\columnwidth]{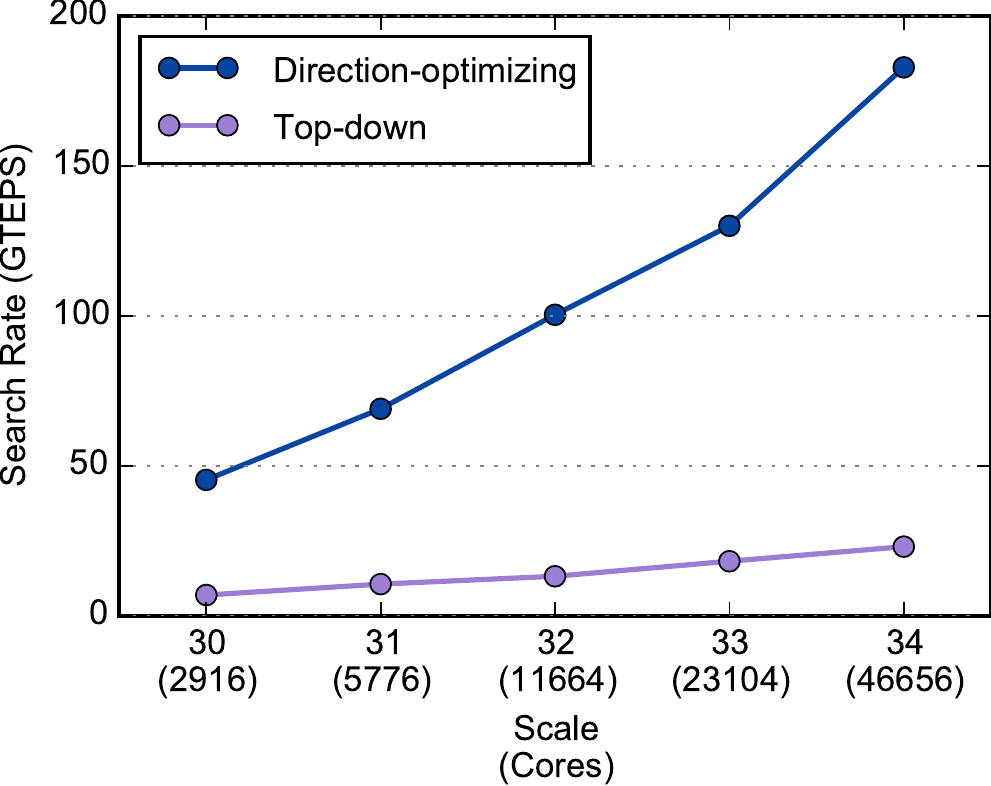}
\caption{R-MAT weak scaling on Titan}
\label{figure:weak-jag}
\end{figure}

Weak scaling results from Titan demonstrate the performance improvement the direction-optimizing implementation gets from the addition of the bottom-up approach (Figure~\ref{figure:weak-jag}). At scale 30 the improvement is 6.5$\times$, but as the graph and system size grow, the ratio of improvements extends to 7.9$\times$. Both implementations are DCSC based. 

\begin{figure}
\centering
\includegraphics[width=0.50\columnwidth]{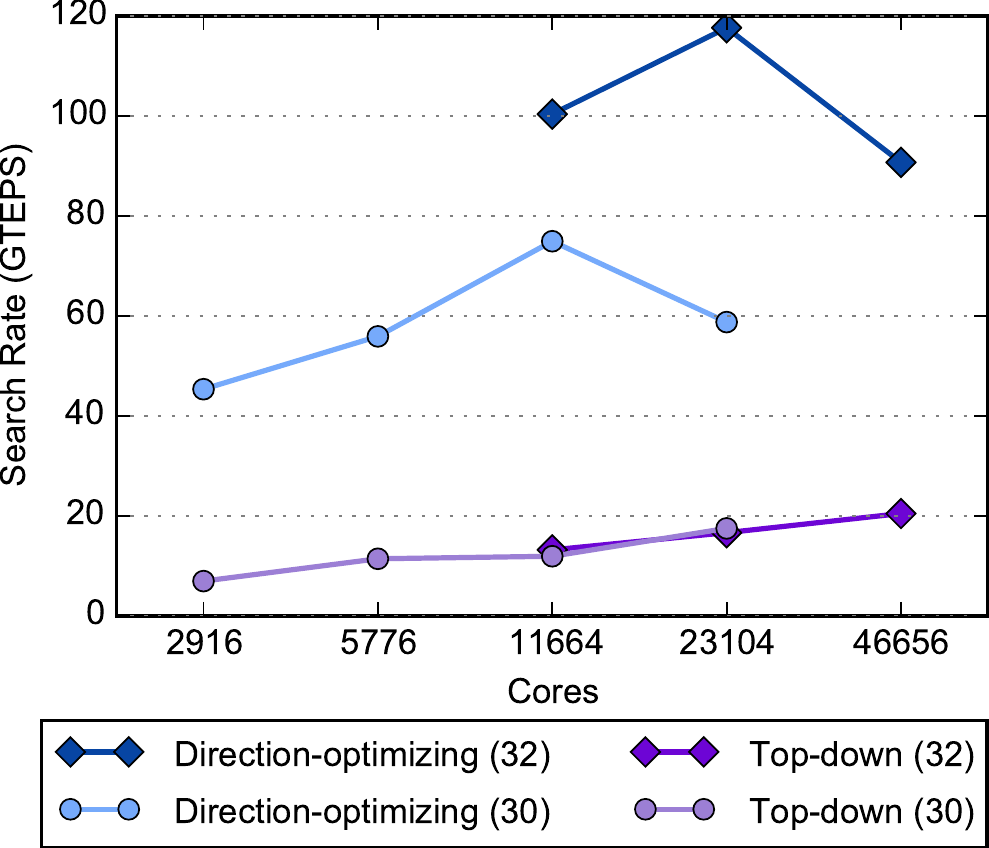}
\caption{R-MAT strong scaling on Titan for graphs of scale 30 and scale 32}
\label{figure:strong-jag}
\end{figure}

Strong scaling results from Titan show promising speedups for the direction-optimizing approach, but it does have some slowdown for a cluster sufficiently large relative to the graph. This behavior is shown on two scales of graph (Figure~\ref{figure:strong-jag}). For both BFS approaches (top-down and bottom-up), increasing the cluster size does have the benefit of reducing the amount of computation per processor, but it comes at the cost of increased communication. The top-down approach does more computation than the bottom-up approach, so this increase in communication is offset by the decrease in computation, producing a net speedup. The bottom-up approach derives some benefit from a larger cluster, but after a certain point the communication overhead hurts its overall performance. Even though it sends less data than the top-down approach, it also has less computation to hide it with. In spite of this slowdown, the direction-optimizing approach still maintains a considerable advantage over the purely top-down implementation. 

\subsection{Platform Performance Comparison}
\begin{figure}[t]
\centering
\includegraphics[width=0.50\columnwidth]{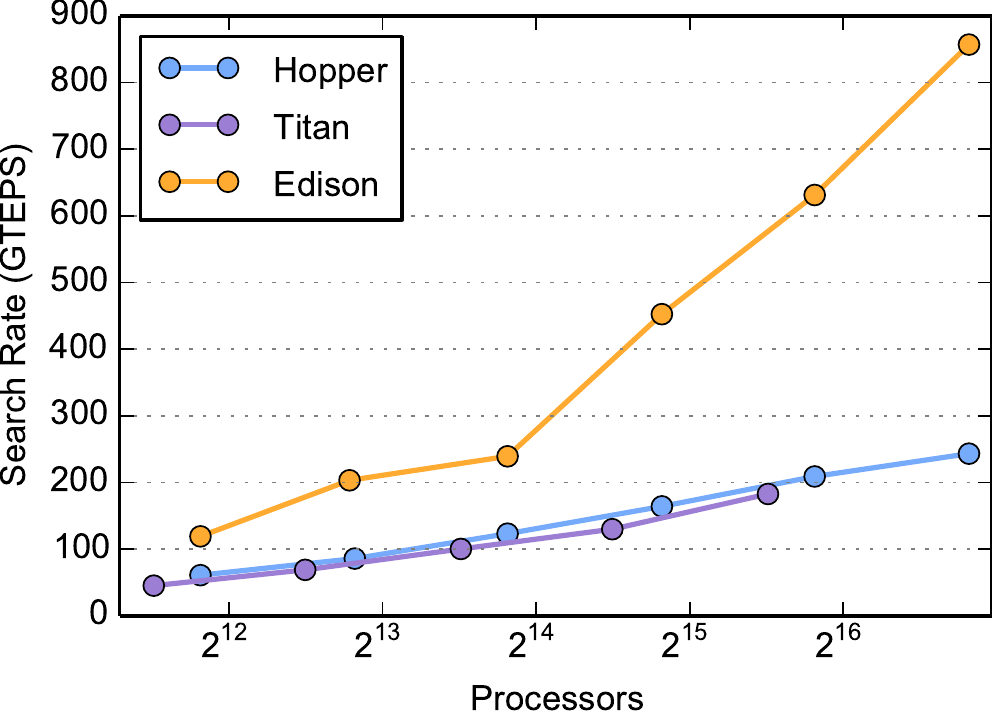}
\caption{R-MAT weak scaling comparison among Hopper, Titan, and Edison. Hopper results start with 3600 cores and Titan start with 2916 cores for scale 30. Due to its higher memory per core,
Edison results start with 3600 cores for scale 31. The number of processors doubles for each increase in scale. In order to smooth out any system congestion effects, the 
runs on Edison has been repeated three times on different dates and we report the harmonic mean of the harmonic means. The largest runs used 115600 cores on Hopper and Edison.}
\label{figure:weak-all}
\end{figure}

The direction-optimizing algorithm shows good weak scaling on all architectures (Figure~\ref{figure:weak-all}). At these cluster and problem sizes, there is no slowdown in performance improvement, indicating a larger allotment on these systems could produce even faster search rates on larger problems. The per-core performance benefit of the Aries and Intel powered XC30 architecture, compared to earlier Gemini and AMD powered XE6 and XK7 architectures is clear. 

\subsection{DCSC vs CSR as Local Data Structures}

\begin{figure}[t]
\centering
\includegraphics[width=0.5 \columnwidth]{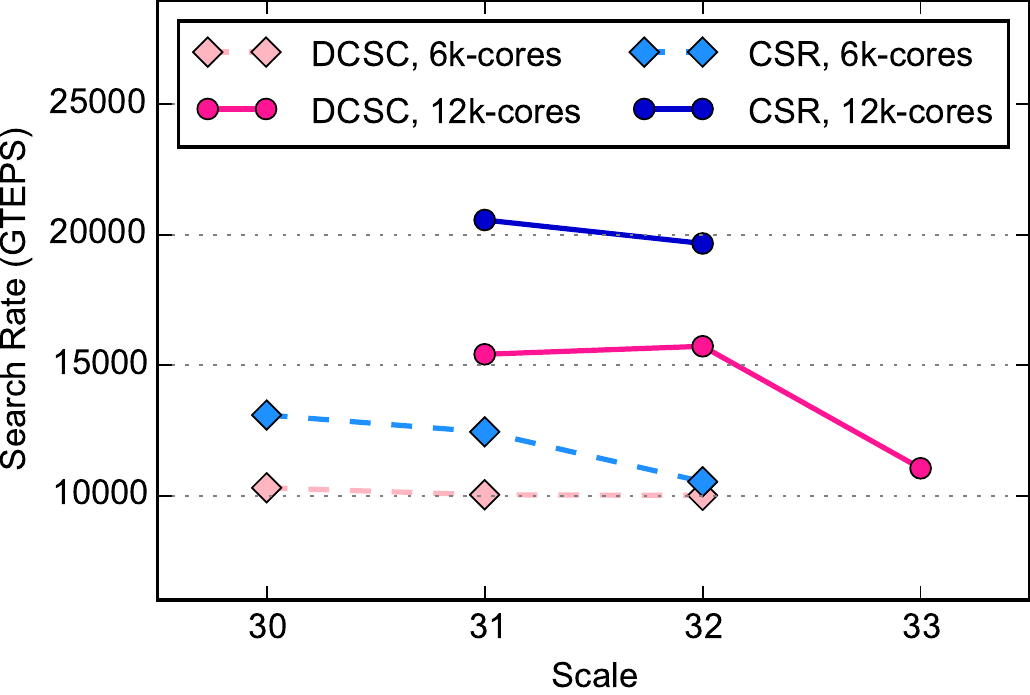}
\caption{DCSC vs CSR comparison on Hopper. The 6k runs are performed on a process grid of $78$x$78$ and the 12k runs are performed on a $120$x$120$ grid.
The x-axis is the size (scale) of the R-MAT graph. }
\label{fig:dcsc-csr}
\end{figure}

In this section, we attempt to capture the relative performance of using CSR or DCSC as local data structure. Our findings are summarized in Figure~\ref{fig:dcsc-csr}.
When the graph fits comfortable into memory, CSR is faster. This is because DCSC has the additional level of indirection when accessing the adjacency of a vertex, 
where the algorithm first has to find the correct vertex (column) pointer by checking the vertex (column) identifiers in its parallel array. The fast access of CSR comes at
the cost of maintaining an $O(n)$ vertex pointers array, which in turns translates into degraded performance as the graph size increases. 

As the graph scale increases, the CSR performance degrades continuously for both the 6k and the 12k runs. The 12k CSR run on the scale 33 ran out of memory 
during graph construction. By contrast, DCSC is slower but more memory efficient. We recommend using a CSR implementation if targeting a fast computation but using
a DCSC implementation if scalability and memory footprint for large concurrencies are important considerations. 

\subsection{Effects of Multithreading}
\begin{figure}[t]
\centering
\includegraphics[width=0.5 \columnwidth]{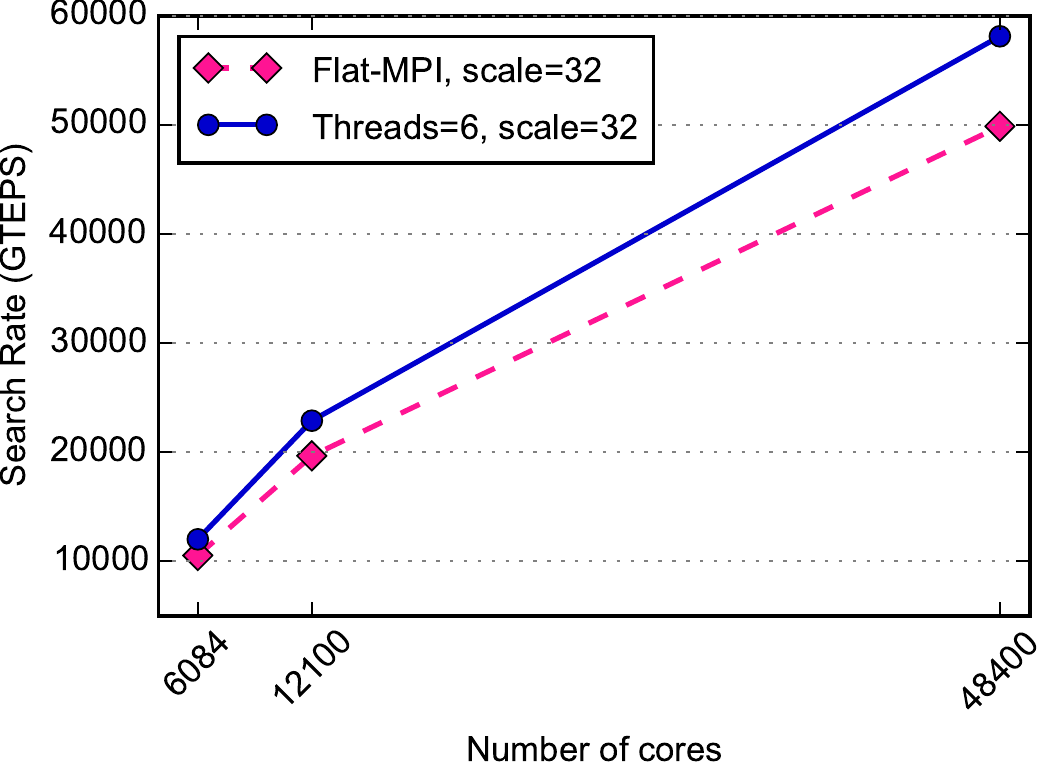}
\caption{Effect of multithreading within the NUMA domain on strong scaling. The experiments use the CSR implementation of the top-down algorithm on Hopper using perfect square grids. Displayed on the 
x-axis are the number of cores used for multithreaded (threads=6) runs. The exact number of cores used for Flat-MPI runs (threads=1) are the closest perfect square to those. }
\label{fig:threadornot}
\end{figure}

Multithreading within NUMA domains are known to reduce memory usage of applications relative to those that map one MPI process per core~\cite{rabenseifner2009hybrid}.
Following our performance model that is described in Section~\ref{sec:anal}, it also reduces the number of participating processors in communication steps, with favorable 
performance implications. In this section, we quantify the effect of multithreading within each NUMA domain using the CSR based implementation of the top-down algorithm.
Our main results is shown in Figure~\ref{fig:threadornot}, where we see a consistent 15-17\% performance improvement for the multithreaded runs compared to the flat-MPI runs. 
 
Not shown in the figure is the better memory scalability of the multithreaded implementation. This is partially due to the decreased cost of maintaining MPI per process information, but also partially 
due to the reduced decremental effect of the CSR scalability issues on 2D decomposition (as the process grid is smaller in the multithreaded case). In particular, several large experiments that successfully 
complete with the multithreaded implementation run out of memory with the flat-MPI implementation: scale 33 using 48k cores with the ``short fat'' grid is one such example.

\subsection{Processor Grid Skewness Impact}
Instead of using a perfect square grid, we can run on rectangular grids. The increase in the skewness of the grid has effects on both communication and computation costs
as they change both the number of processors in collective operations and the sizes of the local data structures. The effects are both predicted by the analysis in 
Section~\ref{sec:anal}. Here, we use the CSR based implementation of the top-down algorithm to analyze the effect of process grid skewness on performance. 
As mentioned in Section~\ref{sec:distimpl}, the CSR based implementation performs Alltoallv along processor columns and Allgatherv along processor rows.
We analyze not only the aggregate performance, but also the changes in communication (for Allgatherv and Alltoallv separately) and computation costs. 

\begin{figure}[!h]
\begin{subfigure}[scale 32]{\includegraphics[width=0.50\linewidth]{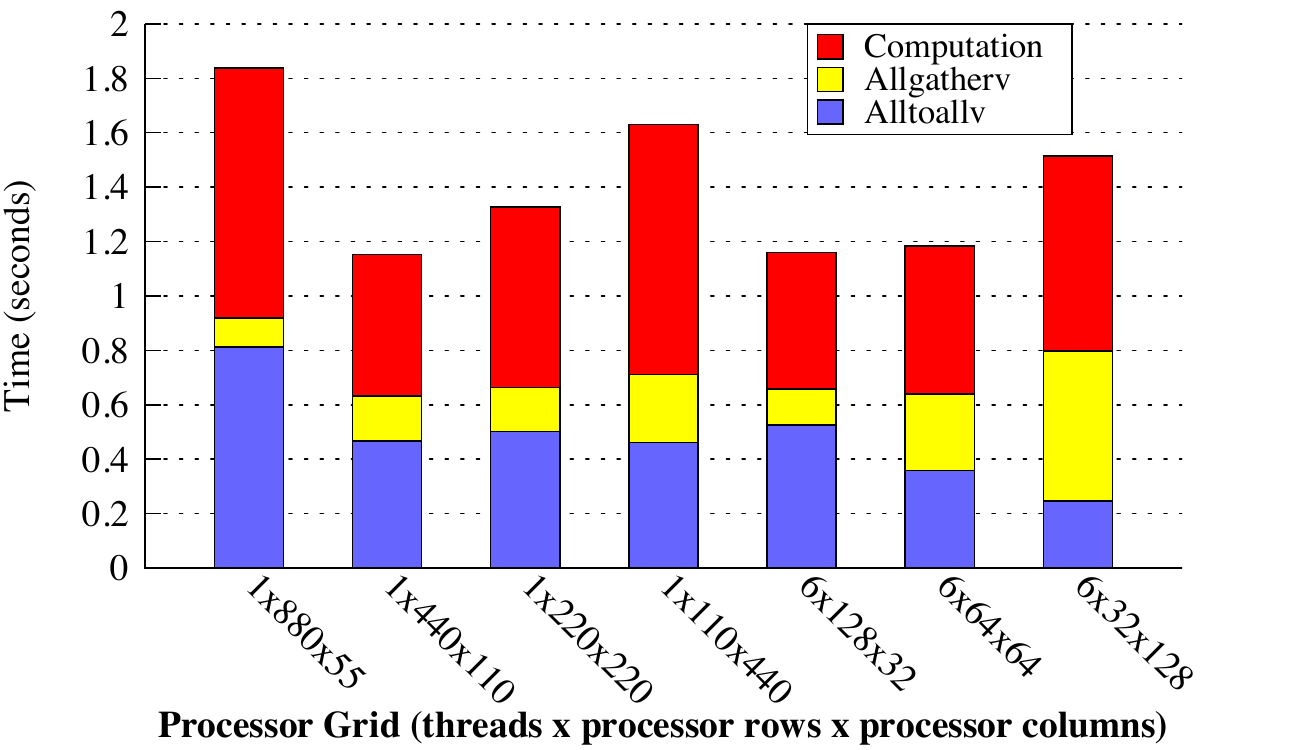}
\label{figure:skew32}}
  \end{subfigure} \hspace{-20pt}
\begin{subfigure}[scale 33]{\includegraphics[width=0.50\linewidth]{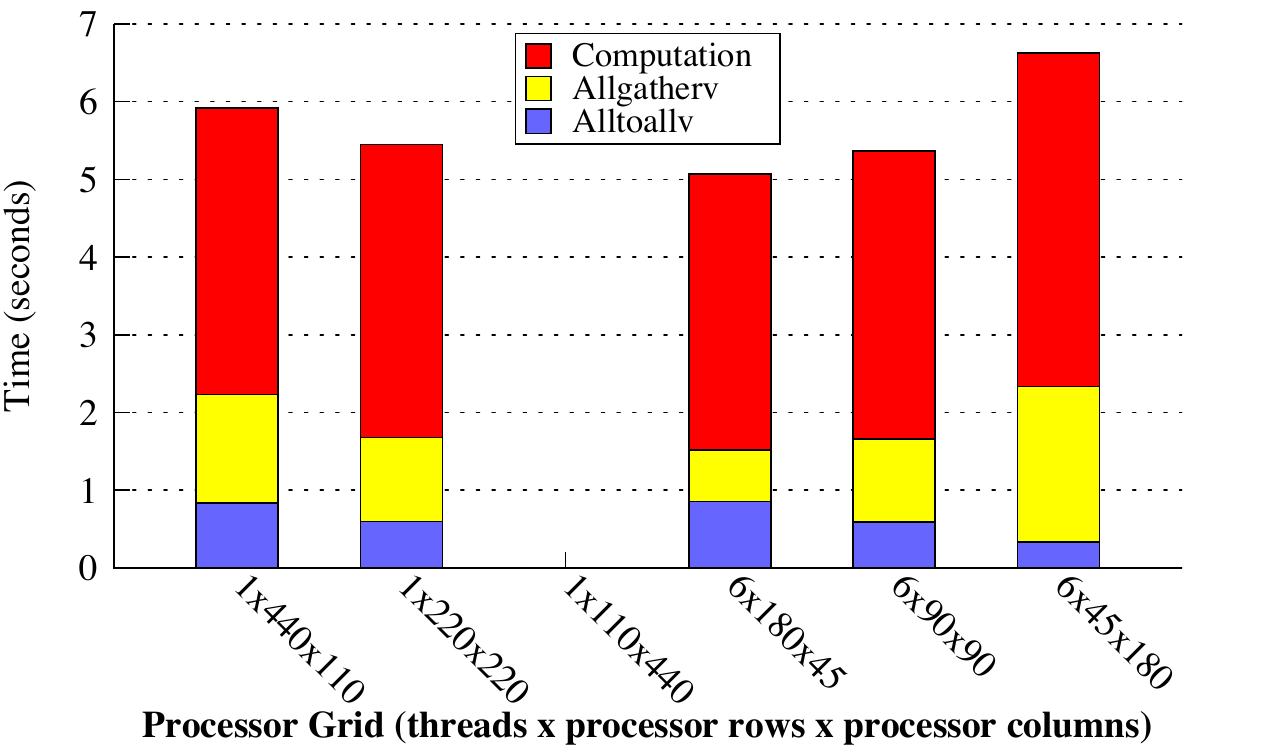}
\label{figure:skew33}}
  \end{subfigure}
  \caption{The effect of process grid skewness for the CSR based implementation of the top-down BFS algorithm. The experiments are run on Hopper at NERSC. 
  Single threaded runs
  used 48400 cores and the multithreaded runs used 48600 cores.
  The $1$x$110$x$440$ configuration, i.e. flat MPI (single threaded) on $110$x$440$ processor grid, ran out of memory for the scale 33 input. }
  \label{figure:skews}
\end{figure}

The results for scale 32 inputs, shown in Figure~\ref{figure:skew32}, favor slightly ``tall skinny'' grids such as $1$x$440$x$110$ and $6$x$180$x$45$, with the square $1$x$220$x$220$ and $6$x$90$x$90$ grids coming a close second. Especially for the multithreaded case, the difference is negligible. The results for scale 33 inputs,
shown in Figure~\ref{figure:skew33}, are less clear. In the single threaded case, the square $220$x$220$ is the fastest whereas in the multithreaded case, the tall skinny grid
of $180$x$45$ is the fastest. In all cases, the ``short fat'' grids performed worse than square and  slightly ``tall skinny'' grids. In one case, the single threaded run on scale 33 even ran out of memory. Even though ``short fat'' grids can often reduce Alltoallv costs by limiting the number of participating processors, especially in the multithreaded case, that is not enough to overcome the increased costs of Allgatherv and local communication. Given that severely skewed ``tall skinny'' grid of $1$x$880$x$55$ performed worst, we recommend using either square grids or slightly ``tall skinny'' grids.

Figure~\ref{figure:skews} uncovers a scalability issue with the CSR based implementation. As evidenced by the out-of-memory condition of the $1$x$110$x$440$ configuration,
scale 33 pushes the implementation close to its memory limits on this architectures. Possibly for that reason, we see $4$X increase in time when going from
scale 32 to scale 33. Most of the performance degradation is due to increased local computation, which provides empirical evidence on the well-document 
CSR scalability issues~\cite{ipdps08} on 2D processor grids. The communication costs perform closer to expected, with approximately $2.5$X increase 
in time when doubling the input size.

\subsection{Strong Scaling on Real Data}

\begin{figure}
\centering
\includegraphics[width=0.5 \columnwidth]{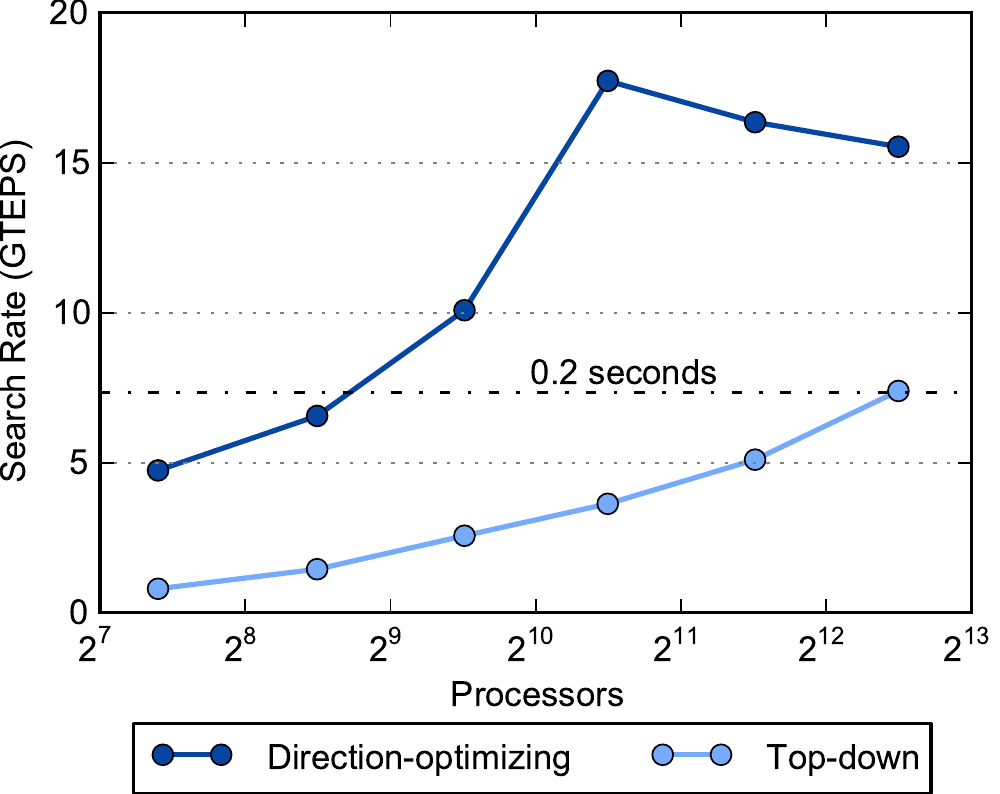}
\caption{Twitter dataset~\cite{Twitter} on Titan}
\label{figure:twitter}
\end{figure}

We also run our implementations on the Twitter dataset~\cite{Twitter} to demonstrate it works well on other scale-free low-diameter graphs (Figure~\ref{figure:twitter}). Because the real-world graph is much smaller than the other graphs in this study, and the direction-optimizing approach already takes only $0.08$ seconds to do a full BFS on this graph with 1440 cores, its performance does not increase any further by
increasing core counts. The direction-optimizing algorithm, however, provides an economic response to the inverse question ``how many core are needed to traverse this data set in less than 0.2 seconds?'', as it requires an order of magnitude less cores for the same performance.

\section*{Acknowledgments}
Preliminary sections of this paper partially appeared in earlier conference proceedings~\cite{beamerbfs13, sc11-bfs2d}.

This material is partially based upon work supported by the U.S.
Department of Energy, Office of Science, Office of Advanced
Scientific Computing Research, Applied Mathematics program
under contract number No. DE-AC02-05CH11231.
Research supported by Microsoft (Award \#024263) and Intel (Award \#024894) funding and by matching funding by U.C. Discovery (Award \#DIG07-10227). 
Additional support comes from Par Lab affiliates National Instruments, Nokia, NVIDIA, Oracle, and Samsung. Partially funded by DARPA Award HR0011-11-C-0100.
This research used resources of the National Energy Research Scientific Computing Center, a DOE Office of Science User Facility supported by the Office of Science of the U.S. Department of Energy under Contract No. DE-AC02-05CH11231.

This research was also supported by the National Science Foundation grant ACI-1253881.

\bibliography{chapter} 

\end{document}